\documentclass[11pt]{article}
\begin{document}
\title{Non-entropic theory of rubber elasticity:\\
flexible chains grafted on a rigid surface}

\author{A.D. Drozdov\footnote{
Phone: 972-86472146.
E-mail: aleksey@bgumail.bgu.ac.il}\\
Department of Chemical Engineering\\
Ben-Gurion University of the Negev\\
P.O. Box 653\\
Beer-Sheva 84105, Israel}
\date{}
\maketitle

\begin{abstract}
The elastic response is studied of (i)
a single flexible chain grafted on a rigid plane
and (ii) an ensemble of non-interacting tethered chains.
It is demonstrated that the entropic theory of
rubber elasticity leads to conclusions that
disagree with experimental data.
A modification of the conventional approach is proposed,
where the end-to-end distribution function
(treated as the governing parameter) is replaced by the
average energy of a chain.
It is revealed that this refinement ensures an adequate
description of the mechanical behavior of flexible
chains.
Results of numerical simulation are compared with
observations on uniaxial compression of a layer of
grafted chains, and an acceptable agreement is shown
between the model predictions and the experimental data.
Based on the analysis of combined compression and shear,
a novel micro-mechanism is proposed for the reduction
of friction of polymer melts at rigid walls.
\end{abstract}
\vspace*{5 mm}

\noindent
{\bf Key-words:}
Flexible chain,
Grafted layer,
Entropic elasticity,
Free energy,
Path integral

\section{Introduction}

This study is concerned with nonlinear elasticity
of an individual flexible chain and of an ensemble of
flexible chains grafted on a rigid surface.
This subject has attracted substantial attention in the
past decade, which may be explained by the importance
of the mechanical responses of grafted polymer chains
and their ensembles in a number of engineering applications,
ranging from
stick-slip transitions in polymer melts near rigid walls
at extrusion \cite{BDeG92,DM03}
to drag reduction in dilute polymer solutions
(reflecting hydrodynamically induced transition from
coiled to stretched conformations of chains \cite{RZ03}),
stabilization of colloidal dispersions by polymers
\cite{RSS89},
enhancement of adhesion by polymer brushes \cite{SGS01},
and manipulation on polymer membranes \cite{HRM01}
and single DNA molecules tethered at surfaces \cite{SB98,FB03,CLA03}.

We assume the surface to be sufficiently smooth in the
sense that its radius of curvature substantially exceeds
the mean square end-to-end distance of a chain $b$.
Under this hypothesis, a tethered macromolecule is thought
of as a chain whose end is fixed at some point on a rigid
plane,
whereas all other points are located in a half-space.
For definiteness, we suppose that the chain lies in the
half-space $X_{3}\geq 0$,
where a Cartesian coordinate frame $\{ X_{m} \}$
with base vectors ${\bf e}_{m}$ ($m=1,2,3$)
is chosen in such a way that the vectors ${\bf e}_{1}$ and
${\bf e}_{2}$ are located in the plane, and ${\bf e}_{3}$
is orthogonal to the plane.

Two models are conventionally employed to describe
conformations of a polymer chain.
According to the first, a chain is treated as a random
walk with $N\gg 1$ steps that starts at the origin.
Assuming the length of a step $b_{0}$ to be constant
and small (in the sense that the contour length $L=b_{0}N$
is finite when $N\to \infty$), we treat the number of
steps $n$ as a continuous variable.
For a flexible chain, the probability density $P(n,{\bf r})$
that the $n$th step ends at a point ${\bf r}$ is governed
by the diffusion equation
\begin{equation}
\frac{\partial P}{\partial n}= \frac{b^{2}}{6} \Delta P,
\end{equation}
where $\Delta$ stands for the Laplace operator,
and $b=\sqrt{b_{0}L}$ \cite{DE86}.
For a Gaussian chain (no restrictions on the walk),
the function $P(n,{\bf r})$ obeys Eq. (1) with
the initial condition
\begin{equation}
P(0,{\bf r})=\delta ({\bf r}),
\end{equation}
where $\delta$ stands for the Dirac delta-function,
and the normalization condition
\begin{equation}
\int P(n,{\bf r})d{\bf r}=1,
\end{equation}
where integration is performed over the entire
space.
For a grafted chain, the function $P(n,{\bf r})$
satisfies the additional adsorbing boundary condition
\cite{diM65}
\begin{equation}
P(n,{\bf r})\biggl |_{X_{3}=0}=0.
\end{equation}
The walk is entirely characterized by the distribution
function $p({\bf Q})=P(N,{\bf Q})$
of the end-to-end vector ${\bf Q}$.
It follows from Eqs. (1) to (4) that $p({\bf Q})$
is given by
\begin{equation}
p_{G}({\bf Q})=\Bigl (\frac{3}{2\pi b^{2}}\Bigr )^{\frac{3}{2}}
\exp \Bigl (-\frac{3Q^{2}}{2b^{2}}\Bigr )
\end{equation}
for a Gaussian chain, and it reads
\begin{equation}
p({\bf Q})=\frac{9Q_{3}}{2\pi b^{4}}
\exp \Bigl (-\frac{3Q^{2}}{2b^{2}}\Bigr )
\end{equation}
for a chain confined to the half-space.
Here $Q_{m}$ is the $m$th Cartesian
coordinate of the vector ${\bf Q}$, $Q=|{\bf Q}|$,
and the subscript index ``G" stands for ``Gaussian."

According to the other approach,
a chain is treated as a curve (with ``length" $L$)
in a three-dimensional space.
An arbitrary configuration of the chain is determined by
the function ${\bf r}(s)$,
where ${\bf r}$ stands for the radius vector,
and $s\in [0,L]$.
For a chain that begins at the origin and finishes
at a point ${\bf Q}$, the function ${\bf r}(s)$
satisfies the boundary conditions
\begin{equation}
{\bf r}(0)={\bf 0},
\qquad
{\bf r}(L)={\bf Q}.
\end{equation}
The energy of a chain is described by a Hamiltonian
$H({\bf r})$, which determines the Green function
(propagator) for the chain
\begin{equation}
G({\bf Q})=\int_{{\bf r}(0)={\bf 0}}^{{\bf r}(L)={\bf Q}}
\exp \biggl [-\frac{H({\bf r}(s))}{k_{\rm B}T}\biggr ]
{\cal D} [{\bf r}(s)],
\end{equation}
where $k_{B}$ is Boltzmann's constant,
$T$ is the absolute temperature,
and the path integral with the measure ${\cal D}[{\bf r}]$
is calculated over all curves ${\bf r}(s)$ that obey Eq. (7).
For a discussion of properties of path integrals,
the reader is referred to \cite{FH65,Kle95}.
As the functional integral is determined up to an arbitrary
multiplier, an additional constraint is imposed on
the function $G({\bf Q})$,
\begin{equation}
\int G({\bf Q}) d{\bf Q}=1 ,
\end{equation}
which allows the Green function $G({\bf Q})$
to be referred to as the distribution function
of end-to-end vectors $p({\bf Q})$.

For a Gaussian chain, the Hamiltonian in Eq. (8) reads
\begin{equation}
H_{G}({\bf r})=\frac{3k_{\rm B}T}{2b_{0}}\int_{0}^{L}
\Bigl (\frac{d{\bf r}}{ds}(s)\Bigr )^{2} ds ,
\end{equation}
and simple algebra implies that the normalized Green
function $G_{G}({\bf Q})$ is given by Eq. (5).
To account for the effect of rigid plane on chain
conformations, a penalty functional is inserted into
the Hamiltonian,
\begin{equation}
H({\bf r})=H_{G}({\bf r})+\frac{A}{L}\int_{0}^{L}
\delta (r_{3}(s))ds,
\end{equation}
where $A$ is a sufficiently large constant,
and $r_{m}(s)$ is the $m$th coordinate of the
vector ${\bf r}(s)$.
The physical meaning of the integral on the right-hand
side of Eq. (11) is the ``number" of times that the chain
crosses the rigid plane in the configuration ${\bf r}(s)$.
When $A\gg 1$, it is natural to expect that the configurations
where the chain crosses the plane have a negligible weight
in the path integral, and they are automatically excluded
from the consideration.
A rigorous analysis \cite{Slu04} demonstrates that this
assertion is true:
for an arbitrary $A>0$, the non-normalized Green function
reads
\begin{equation}
G(a,{\bf Q}) = 1-2 \int_{0}^{1} \exp \Bigl (
-\frac{3Q_{3}^{2}\tau^{2}}{2b^{2}(1-\tau^{2})}
\Bigr ) \varphi(a\tau)\frac{d\tau}{\tau\sqrt{1-\tau^{2}}}
\end{equation}
with
\begin{equation}
a=\frac{A}{k_{B}Tb}\sqrt{\frac{3}{2}},
\qquad
\varphi(x) = \frac{x}{\sqrt{\pi}}-x^{2}\exp (x^{2}){\rm erfc}(x),
\end{equation}
where ${\rm erfc}(x)$ is the complement error function.
It is easy to show that after taking the limit $A\to\infty$
and normalization, the Green function (12) coincides with
the function $p({\bf Q})$ given by Eq. (6).

The classical theory of entropic elasticity has been
formulated more than half a century ago, and its
exposition can be found in a number of monographs,
see, e.g., \cite{Tre75,Fer80}.
This concept is grounded on the Boltzmann formula
that expresses the probability density $p({\bf Q})$
in terms of the free energy of a chain $\Psi({\bf Q})$,
\begin{equation}
p({\bf Q})=\exp \Bigl (-\frac{\Psi({\bf Q})}{k_{B}T}\Bigr ).
\end{equation}
According to the finite elasticity theory,
two states of a chain are distinguished:
(i) the reference (initial) state occupied
before application of external loads,
and (ii) the actual (deformed) state that the chain
acquires after deformation.
The end-to-end vector of a chain in the deformed
state ${\bf Q}^{\prime}$ is expressed in terms
of that in the reference state ${\bf Q}$ by the formula
\begin{equation}
{\bf Q}^{\prime}={\bf F}\cdot {\bf Q},
\end{equation}
where ${\bf F}$ is the deformation gradient,
and the dot stands for inner product
(for simplicity, we adopt the affinity hypothesis,
according to which, the deformation gradient at
the micro-level coincides with the deformation
gradient for macro-deformation).
It follows from Eqs. (14) and (15) that the increment of
free energy $\Delta \Psi({\bf F},{\bf Q})$
driven by deformation of a chain reads
\[
\Delta \Psi ({\bf F},{\bf Q})=-k_{B}T \Bigl [
\ln p({\bf F}\cdot {\bf Q})-\ln p ({\bf Q}) \Bigr ].
\]
The strain energy per chain $W({\bf F})$ is
calculated by averaging the increment of free energy
over the initial distribution of end-to-end vectors,
\begin{equation}
W({\bf F})=k_{B}T \int \Bigl [
\ln p ({\bf Q})-\ln p({\bf F}\cdot {\bf Q})\Bigr ]
p({\bf Q}) d{\bf Q},
\end{equation}
where the integration is performed over the entire space.
Given a strain energy $W({\bf F})$, constitutive equations for
a chain or an ensemble of chains (whose strain
energy equals the sum of strain energies of individual chains)
can be determined by conventional formulas.
For example, for a Gaussian chain with the distribution
function (5), Eq. (16) implies the classical formula
\[
W=\frac{1}{2} k_{B}T \Bigl ({\cal I}_{1} ({\bf C})-3 \Bigr ),
\]
where ${\cal I}_{m}$ stands for the $m$th principal invariant
of a tensor,
${\bf C}={\bf F}^{\top}\cdot {\bf F}$ is the right
Cauchy-Green deformation tensor,
and $\top$ denotes transpose.
For a chain grafted on a rigid plane, the natural restriction
is imposed on the deformation gradient ${\bf F}$:
the end-to-end vectors before and after deformation,
${\bf Q}$ and ${\bf Q}^{\prime}$,
are located in the same half-space.
Although this constraint complicates calculations
in the general case, stress--strain relations can be
easily developed in an explicit form for deformation
programs traditionally studied in nonlinear mechanics.

The starting point of the present work is the difference
between the two models for polymer chains.
For a chain modeled as a random walk, the Boltzmann
equation (14) provides the only way to introduce free energy.
On the contrary, for a chain described as a curve with
a Hamiltonian $H$ ascribed to it, the free energy
$\Psi({\bf Q})$ may be identified as the average value
of the Hamiltonian
$\langle H\rangle_{\bf Q}$,
\begin{equation}
\Psi({\bf Q})=\frac{1}{G({\bf Q})}
\int_{{\bf r}(0)={\bf 0}}^{{\bf r}(L)={\bf Q}}
H({\bf r}(s)) \exp \biggl [-\frac{H({\bf r}(s))}{k_{\rm B}T}
\biggr ] {\cal D} [{\bf r}(s)],
\end{equation}
where the pre-factor $1/G({\bf Q})$ plays the role of
the partition function in the conventional formula
for averaging.
Given a free energy $\Psi({\bf Q})$, we can define
the non-normalized distribution function of
end-to-end vectors $p_{\Psi}({\bf Q})$ by Eq. (14),
and, after appropriate normalization of $p_{\Psi}({\bf Q})$
with the help of Eq. (3),
calculate the strain energy per chain $W({\bf F})$
by the formula similar to Eq. (16),
\begin{equation}
W({\bf F})=\int \Bigl [ \Psi({\bf F}\cdot {\bf Q})
-\Psi({\bf Q})\Bigr ] p_{\Psi}({\bf Q}) d{\bf Q}.
\end{equation}
Three questions arise naturally after the formulation
of this ``upside-down" (with respect to the entropic
elasticity theory) approach to the derivation of
constitutive equations: (i) does this technique lead
to classical results for a Gaussian chain,
(ii) is there any case of practical interest where
the novel method implies physically plausible
conclusions, whereas the traditional approach fails,
and (iii) how difficult is it to calculate the path
integral in Eq. (17) provided that the Green function
$G({\bf Q})$ is known?
The objective of this paper is to shed some light
on these issues.

The exposition is organized as follows.
In Section 2, a method is described that allows
the average free energy $\Psi({\bf Q})$ to be found
from Eq. (17) without calculation of the path
integral.
In particular, it is shown that for a Gaussian
chain, constitutive equations (16) and (18) coincide.
Force--stretch relations for an individual chain
grafted on a rigid surface are developed in
Section 3.
It is demonstrated that two approaches lead to
qualitatively similar conclusions for tension
(compression) of a chain, although the results
differ quantitatively.
Stress--strain relations for an ensemble of
noninteracting flexible chains grafted on a
rigid surface are developed in Section 4.
It is revealed that the conventional
theory of rubber elasticity fails to
adequately describe observations, whereas
the non-entropic concept provides good agreement
with experimental data.
Combined shear and compression of a layer
of non-interacting grafted chains is studied in
Section 5, where a novel mechanism is suggested
for the reduction of friction of polymer melts
near rigid surfaces (in addition to the
standard entanglement--disentanglement process).
Some concluding remarks are formulated in
Section 6.

\section{Calculation of the free energy}

Our aim is to determine the function $\Psi({\bf Q})$
from Eq. (17) explicitly, provided that the Green
function $G({\bf Q})$ for a flexible chain is known.
To simplify transformations, it is convenient to consider
a more general problem of averaging for an arbitrary
Hamiltonian $H$ of the form
\begin{equation}
H({\bf r})=H_{G}({\bf r})+A\Phi({\bf r}),
\end{equation}
where $\Phi({\bf r})$ is an arbitrary potential
that describes intra-chain and inter-chain
interactions.
According to Eqs. (8) and (10), the Green function for
the Hamiltonian (19) reads
\begin{equation}
G_{H}(A,b_{0},{\bf Q})
=\int_{{\bf r}(0)={\bf 0}}^{{\bf r}(L)={\bf Q}}
\exp \biggl [-\frac{1}{k_{\rm B}T}
\Bigl (\frac{3k_{\rm B}T}{2b_{0}}\int_{0}^{L}
\Bigl (\frac{d{\bf r}}{ds}(s)\Bigr )^{2} ds
+A\Phi({\bf r}(s))\Bigr )\biggr ] {\cal D} [{\bf r}(s)],
\end{equation}
where the parameters $b_{0}$ and $A$ are included explicitly
as arguments of the function $G_{H}$.
Differentiating Eq. (20) with respect to $b_{0}$ and
using Eqs. (10) and (17), we find that
\begin{eqnarray}
\frac{\partial G_{H}}{\partial b_{0}}
&=& \frac{1}{k_{B}T b_{0}}
\int_{{\bf r}(0)={\bf 0}}^{{\bf r}(L)={\bf Q}}
H_{G}({\bf r}(s)) \exp \biggl [-\frac{1}{k_{\rm B}T}
\Bigl (\frac{3k_{\rm B}T}{2b_{0}}\int_{0}^{L}
\Bigl (\frac{d{\bf r}}{ds}(s)\Bigr )^{2} ds
+A\Phi({\bf r}(s))\Bigr )\biggr ] {\cal D} [{\bf r}(s)]
\nonumber\\
&=& \frac{1}{k_{B}T b_{0}} \langle H_{G}\rangle_{\bf Q}
G_{H}.
\end{eqnarray}
Similarly, differentiation of Eq. (20) with respect to $A$
results in
\begin{equation}
\frac{\partial G_{H}}{\partial A}
=-\frac{1}{k_{B}T} \langle \Phi\rangle_{\bf Q} G_{H}.
\end{equation}
It follows from Eqs. (17), (21) and (22) that
\begin{eqnarray}
\Psi({\bf Q}) &=& \langle H_{G}+A\Phi\rangle_{\bf Q}
\nonumber\\
&=& \frac{k_{B}T}{G_{H}(A,b_{0},{\bf Q})}
\biggl [b_{0}
\frac{\partial G_{H}}{\partial b_{0}}(A,b_{0},{\bf Q})
-A\frac{\partial G_{H}}{\partial A}(A,b_{0},{\bf Q})
\biggr ].
\end{eqnarray}
Formula (23) provides an analytical expression for
the average free energy of a flexible chain with
an arbitrary potential $\Phi$.
For example, for a Gaussian chain ($A=0$),
Eqs. (5) and (23) imply that
\begin{equation}
\Psi_{G}({\bf Q})=\frac{3k_{B}T}{2}
\Bigl (\frac{Q^{2}}{b^{2}}-1 \Bigr ).
\end{equation}
Evidently, insertion of expression (24) into Eq. (14)
and subsequent normalization results in the distribution
function given by Eq. (5), which means that
for a Gaussian chain, our approach coincides with
the conventional one.

For a flexible chain confined to a half-space,
we calculate the averages $\langle H_{G}\rangle_{\bf Q}$
and $\langle A\Phi\rangle_{\bf Q}$ separately
and use the first equality in Eq. (23).
Equation (21) results in
\[
\langle H_{G}\rangle_{\bf Q}=\frac{k_{B}Tb_{0}}{G_{H}}
\frac{\partial G_{H}}{\partial b_{0}}.
\]
As the differentiation with respect to $b_{0}$
is independent of $A$, we can use Eq. (6) for the
limit of the Green function when $A\to\infty$.
Substitution of Eq. (6) into this equality implies that
\begin{equation}
\langle H_{G}\rangle_{\bf Q}=k_{B}T \Bigl (\frac{3Q^{2}}{2b^{2}}
-2\Bigr ).
\end{equation}
Comparison of Eqs. (24) and (25) shows that
the average Gaussian Hamiltonian is not affected
by the presence of the constraint (additive constants in
these relations do not influence the increment of
free energy $\Delta \Psi$).

It follows from Eqs. (12), (13) and (22) that for any $A>0$,
\begin{equation}
\langle A\Phi\rangle_{\bf Q}=-\frac{k_{B}TA}{G_{H}}
\frac{\partial G_{H}}{\partial A}
=-\frac{k_{B}Ta}{G(a,{\bf Q})}
\frac{\partial G}{\partial a}(a,{\bf Q}),
\end{equation}
where $G(a,{\bf Q})$ is given by Eq. (12).
Differentiation of Eq. (12) with respect to $a$ results in
\[
\frac{\partial G}{\partial a}(a,{\bf Q})
= -2 \int_{0}^{1} \exp \Bigl (
-\frac{3Q_{3}^{2}\tau^{2}}{2b^{2}(1-\tau^{2})}
\Bigr ) \varphi^{\prime}(a\tau)\frac{d\tau}{\sqrt{1-\tau^{2}}},
\]
where the prime denotes differentiation with respect to
the argument.
Bearing in mind that
\[
\frac{\partial \varphi}{\partial \tau}(a\tau)
=a \varphi^{\prime}(a\tau)
\]
and integrating by parts,
we present this equality in the form
\begin{eqnarray}
\frac{\partial G}{\partial a}(a,{\bf Q})
&=& -\frac{2}{a} \biggl [\exp \Bigl (
-\frac{3Q_{3}^{2}\tau^{2}}{2b^{2}(1-\tau^{2})}
\Bigr )\frac{\varphi(a\tau)}{\sqrt{1-\tau^{2}}}
\biggr ]_{\tau=0}^{\tau=1}
\nonumber\\
&& +\frac{2}{a} \int_{0}^{1} \varphi(a\tau)
\frac{\partial}{\partial\tau}
\Bigl [ \exp \Bigl (
-\frac{3Q_{3}^{2}\tau^{2}}{2b^{2}(1-\tau^{2})}
\Bigr )\frac{1}{\sqrt{1-\tau^{2}}}\Bigr ]
d\tau.
\end{eqnarray}
The out-of-integral term vanishes (because of the
property of the exponent at $\tau=1$ and due
to Eq. (13) for the function $\varphi(x)$ at
$\tau=0$).
Insertion of expression (27) into Eq. (26) yields
\[
\lim_{A\to\infty} \langle A\Phi\rangle_{\bf Q}
=-2k_{B}T \frac{
\lim_{a\to\infty} \int_{0}^{1} a\varphi(a\tau)
\frac{\partial}{\partial\tau}
\Bigl [ \exp \Bigl (
-\frac{3Q_{3}^{2}\tau^{2}}{2b^{2}(1-\tau^{2})}
\Bigr )\frac{1}{\sqrt{1-\tau^{2}}}\Bigr ]
d\tau }
{\lim_{a\to\infty} aG(a,{\bf Q})}.
\]
The limit of the function $aG(a,{\bf Q})$ was
calculated in \cite{Slu04},
\[
\lim_{a\to\infty} aG(a,{\bf Q})
=\frac{Q_{3}}{b}\sqrt{\frac{3}{2}}.
\]
Taking into account that
\[
\lim_{a\to \infty} a\varphi(a\tau)=\frac{1}{2\sqrt{\pi}\tau}
\]
and performing differentiation with respect to $\tau$,
we arrive at the formula
\begin{eqnarray*}
\lim_{A\to\infty} \langle A\Phi\rangle_{\bf Q}
&=& \frac{k_{B}Tb}{Q_{3}} \sqrt{\frac{2}{3\pi}}
\int_{0}^{1} \exp \Bigl (
-\frac{3Q_{3}^{2}\tau^{2}}{2b^{2}(1-\tau^{2})}\Bigr )
\Bigl (\frac{3Q_{3}^{2}}{b^{2}(1-\tau^{2})}-1\Bigr )
\frac{d\tau}{(1-\tau^{2})^{\frac{3}{2}}}
\nonumber\\
&=& \frac{k_{B}Tb}{Q_{3}} \sqrt{\frac{2}{3\pi}}
\int_{0}^{\infty} \exp \Bigl (
-\frac{3Q_{3}^{2}}{2b^{2}}t^{2}\Bigr )
\Bigl (\frac{3Q_{3}^{2}}{b^{2}}(1+t^{2})-1\Bigr )dt,
\end{eqnarray*}
where $t=\tau/\sqrt{1-\tau^{2}}$.
Calculation of the Gaussian integral results in
\[
\lim_{A\to\infty} \langle A\Phi\rangle_{\bf Q}
=k_{B}T\Bigl (\frac{b^{2}}{3Q_{3}^{2}}+1\Bigr ),
\]
which, together with Eqs. (23) and (25), implies that
\begin{equation}
\Psi({\bf Q})=k_{B}T \Bigl (\frac{3Q^{2}}{2b^{2}}
+\frac{b^{2}}{3Q_{3}^{2}}-1\Bigr ).
\end{equation}
Formula (28) provides an explicit expression for
the average free energy of a flexible chain grafted
on a rigid surface.

\section{Uniaxial tension of an individual chain}

Our aim now is to compare the mechanical response of
an individual flexible chain grafted on a rigid
plane when its free energy is (i) determined by
the entropic elasticity theory, Eq. (14),
and (ii) given by Eq. (28).

We begin with the conventional approach, substitute
expression (6) into Eq. (14), and find that
\begin{equation}
\Psi_{\rm e}({\bf Q})=k_{B}T\Bigl (\frac{3Q^{2}}{2b^{2}}
-\ln \frac{9 Q_{3}}{2\pi b^{4}} \Bigr ),
\end{equation}
where the subscript index ``e" stands for ``entropic."
To determine the natural state of the chain,
that is the state (described by an
end-to-end vector ${\bf Q}^{0}$) in which the free
energy has its extremum, we differentiate Eq. (29) with
respect to $Q_{m}$, equate the derivatives to zero,
and obtain
\begin{equation}
Q_{1}^{0}=0,
\qquad
Q_{2}^{0}=0,
\qquad
Q_{3}^{0}=\frac{b}{\sqrt{3}}.
\end{equation}
It follows from Eqs. (29) and (30) that for a fixed $Q_{3}$
(shear), the response of a tethered flexible chain
coincides with that of a Gaussian chain.
Thus, we concentrate on tension (compression)
perpendicular to the rigid plane.
Denote by $z$ the displacement (from the natural state)
of the free end of the chain along the ${\bf e}_{3}$
vector when a force ${\bf f}=f{\bf e}_{3}$ is applied
to this end.
Given an elastic energy $\Psi$, the force $f$ is
expressed in terms of the displacement $z$ by the
formula
\begin{equation}
f=\frac{\partial \Psi}{\partial Q_{3}}
\biggl |_{Q_{3}=Q_{3}^{0}+z}.
\end{equation}
Combination of Eqs. (29) to (31) implies that
\begin{equation}
f_{\rm e}(z)=\frac{3k_{B}T}{b}\Bigl [ \frac{z}{b}
+\frac{1}{\sqrt{3}}\Bigl (1-\Bigl (1
+\frac{z}{b} \sqrt{3}\Bigr )^{-1}\Bigr )
\Bigr ].
\end{equation}
The first term on the right-hand side of Eq. (32)
describes the mechanical response of a Gaussian chain,
whereas the last term characterizes the influence
of the constraint on the force--stretch relation.
The expression in the parentheses approaches zero as
$z\to\infty$, which means that at large extensions,
the effect of surface becomes insignificant.
This expression is important at compression ($z<0$),
because it tends to infinity when $z\to -Q_{3}^{0}$ (the
total compression of the chain requires an
infinite force).
The stiffness (an equivalent spring constant) of the chain
$\mu$ is determined as
\begin{equation}
\mu=\frac{df}{dz}\biggl |_{z=0}.
\end{equation}
Equations (32) and (33) imply that
\begin{equation}
\mu_{\rm e}=6k_{B}T b^{-2},
\end{equation}
which means that the presence of a rigid surface
increases the stiffness of a Gaussian chain
$\mu_{G}=3k_{B}T b^{-2}$ \cite{deG92} by twice.

We now repeat the same calculations for a tethered
flexible chain with free energy (28).
Differentiating Eq. (28) with respect to $Q_{m}$
and equating the derivatives to zero, we find
the end-to-end vector in equilibrium
\begin{equation}
Q_{1}^{0}=0,
\qquad
Q_{2}^{0}=0,
\qquad
Q_{3}^{0}=b\sqrt[4]{\frac{2}{9}}.
\end{equation}
Substitution of expressions (28) and (35) into Eq. (31)
implies the force--stretch relation
\begin{equation}
f_{\rm ne}(z)= \frac{3k_{B}T}{b}\Bigl [ \frac{z}{b}
+\sqrt[4]{\frac{2}{9}}\Bigl (1-\Bigl (1
+\frac{z}{b} \sqrt[4]{\frac{2}{9}}\Bigr )^{-3}\Bigr )
\Bigr ],
\end{equation}
where the subscript index ``ne" stands for ``non-entropic."
Equations (33) and (36) result in
\begin{equation}
\mu_{\rm ne}=12k_{B}T b^{-2},
\end{equation}
which means that the stiffness of a flexible chain
grafted on a rigid surface exceeds that of a Gaussian
chain by a factor of four.

Equations (32), (34) and (36), (37) show that,
although the force--extension relations derived
within these approaches are quantitatively
different, they are quite similar qualitatively.
To reveal a qualitative difference between the two
concepts, we study the elastic behavior of an ensemble
of grafted chains.

\section{Uniaxial tension of an ensemble of chains}

We now consider the response of an ensemble of non-interacting
flexible chains grafted on a rigid plane.
In the rubber elasticity theory, inter-chain interactions
are conventionally accounted
for in terms of the incompressibility condition \cite{DE86},
which means that the neglect of interactions between chains
is tantamount to the assumption about compressibility
of the ensemble.
The difference between the analysis of an individual
chain and that of an ensemble of macromolecules is that
we do not assume end-to-end vectors of chains in
an ensemble to be in their natural states, but suppose
that the distribution of end-to-end vectors ${\bf Q}$ is
governed by an appropriate probability density.

At uniaxial tension in the direction orthogonal
to the plane, the deformation gradient ${\bf F}$
reads
\begin{equation}
{\bf F}={\bf e}_{1}{\bf e}_{1}
+{\bf e}_{2}{\bf e}_{2}
+\lambda {\bf e}_{3}{\bf e}_{3},
\end{equation}
where $\lambda$ stands for the elongation ratio.
It follows from Eqs. (6), (14) and (38) that within
the entropic elasticity theory, the increment of free
energy per chain is given by
\[
\Delta \Psi_{\rm e}(\lambda,{\bf Q})=k_{B}T\Bigl (
\frac{3Q_{3}^{2}}{2b^{2}}(\lambda^{2}-1)-\ln \lambda
\Bigr ).
\]
Inserting this expression and Eq. (6) into Eq. (16)
and calculating the integral, we find the strain energy
of an individual chain,
\begin{equation}
W_{\rm e}(\lambda)=k_{B}T (\lambda^{2}-\ln \lambda-1).
\end{equation}
Denote by $\zeta$ concentration of grafted chains
(the number of chains per unit area of the surface).
Multiplying the strain energy of a chain $W$ by the number
of chains per unit area, we determine the strain energy
per unit area of a layer,
\begin{equation}
\tilde{W}=\zeta W.
\end{equation}
At uniaxial extension, the work of external forces (per unit
area and per unit time) reads
\[
\Pi=\Sigma h \frac{1}{\lambda}\frac{d\lambda}{dt},
\]
where $\Sigma$ is the tensile stress,
$\lambda^{-1}d\lambda/dt$ is the rate of
longitudinal strain,
and $h$ is the initial height of the layer.
According to the first law of thermodynamics,
\begin{equation}
\frac{d\tilde{W}}{dt}=\Pi.
\end{equation}
Combining these relations, we arrive at the formula
\begin{equation}
\Sigma=\lambda \frac{\zeta}{h}\frac{dW}{d\lambda}.
\end{equation}
Equation (42) is similar to the well-known
formula that expresses the principal Cauchy stresses
$\Sigma_{m}$ in terms of the principal elongation
ratios $\lambda_{m}$,
\[
\Sigma_{m}=\lambda_{m} \frac{dW}{d\lambda_{m}},
\]
where $W$ is the strain energy per unit volume
in the reference state and $m=1,2,3$.
Inserting expression (39) into Eq. (42), we find that
\begin{equation}
\Sigma_{\rm e}(\lambda) =\frac{\zeta k_{B}T}{h}(2\lambda^{2}-1).
\end{equation}
Before discussing the physical meaning of Eq. (43),
we find a stress-strain relation for a layer of grafted
chain with the average free energy given by Eq. (28).
It follows from Eqs. (28) and (38) that the increment of
free energy $\Delta\Psi_{\rm ne}$ reads
\[
\Delta \Psi_{\rm ne}(\lambda,{\bf Q})=k_{B}T\Bigl [
\frac{3Q_{3}^{2}}{2b^{2}}(\lambda^{2}-1)
+\frac{b^{2}}{3Q_{3}^{2}}
\Bigl (\frac{1}{\lambda^{2}}-1\Bigr )\Bigr ].
\]
The expression for the distribution function of
end-to-end vectors is found from Eqs. (14) and (28),
\begin{equation}
p_{\rm ne}({\bf Q})=C\exp \Bigl [-\Bigl (
\frac{3Q^{2}}{2b^{2}}+\frac{b^{2}}{3Q_{3}^{2}}
\Bigr )\Bigr ],
\end{equation}
where the pre-factor $C$ is determined by the
normalization condition (3).
Calculation of the integrals over $Q_{1}$ and $Q_{2}$
in this relation implies that
\begin{equation}
C=\frac{1}{2\pi} \Bigl (\frac{3}{b^{2}}\Bigr )^{\frac{3}{2}}
C_{0},
\qquad
C_{0}=\Bigl [ \int_{0}^{\infty}\exp
\Bigl (-\Bigl (\frac{1}{2}z^{2}+\frac{1}{z^{2}}\Bigr )
\Bigr )dz \Bigr ]^{-1},
\end{equation}
where $z=Q_{3}\sqrt{3}/b$.
Substitution of these expressions into Eq. (18) results in
\begin{equation}
W_{\rm ne}(\lambda)=k_{B}T\Bigl [ C_{1}(\lambda^{2}-1)
+C_{2}\Bigl (\frac{1}{\lambda^{2}}-1\Bigr )\Bigr ],
\end{equation}
where
\begin{equation}
C_{1}=C_{0}\int_{0}^{\infty}\exp
\Bigl (-\Bigl (\frac{1}{2}z^{2}+\frac{1}{z^{2}}\Bigr )
\Bigr ) \frac{z^{2}}{2}dz,
\qquad
C_{2}=C_{0}\int_{0}^{\infty}\exp
\Bigl (-\Bigl (\frac{1}{2}z^{2}+\frac{1}{z^{2}}\Bigr )
\Bigr )\frac{dz}{z^{2}}.
\end{equation}
It follows from Eqs. (42) and (46) that
\begin{equation}
\Sigma_{\rm ne}(\lambda)=2\frac{\zeta k_{B}T}{h} \Bigl (
C_{1}\lambda^{2}-C_{2}\lambda^{-2}\Bigr ).
\end{equation}
To find the coefficients $C_{1}$ and $C_{2}$ explicitly,
we transform the second equality in Eq. (45) by
integration by parts
\[
\frac{1}{C_{0}}=2 \int_{0}^{\infty}\exp
\Bigl (-\Bigl (\frac{1}{2}z^{2}+\frac{1}{z^{2}}\Bigr )
\Bigr ) \Bigl (\frac{z^{2}}{2}-\frac{1}{z^{2}}\Bigr ) dz.
\]
Combining this relation with Eq. (47), we conclude that
\begin{equation}
C_{1}-C_{2}=\frac{1}{2}.
\end{equation}
We now set $y=\sqrt{2}/z$ in the second equality in
Eq. (45) to obtain
\[
\frac{1}{C_{0}}=\sqrt{2} \int_{0}^{\infty}\exp
\Bigl (-\Bigl (\frac{1}{2}y^{2}+\frac{1}{y^{2}}\Bigr )
\Bigr ) \frac{1}{y^{2}} dy.
\]
It follows from this formula and Eq. (47) that
\begin{equation}
C_{2}=\frac{1}{\sqrt{2}}.
\end{equation}
Equations (48) to (50) yield
\begin{equation}
\Sigma_{\rm ne}(\lambda)=\sqrt{2} \frac{\zeta k_{B}T}{h}
\Bigl [ \Bigl ( 1+\frac{1}{\sqrt{2}}\Bigr )\lambda^{2}
-\lambda^{-2}\Bigr ].
\end{equation}
According to Eqs. (43) and (51), the tensile stress
$\Sigma_{0}=\zeta k_{B}T/h$ should be applied to an
ensemble of grafted chains in order to maintain its
initial (undeformed) state.
Splitting the longitudinal stress $\Sigma$ into the
sum of the residual stress $\Sigma_{0}$ and the
extra-stress $\sigma$, we find that
\begin{eqnarray}
\sigma_{\rm e}(\lambda) &=& 2 \frac{\zeta k_{B}T}{h}(\lambda^{2}-1),
\\
\sigma_{\rm ne}(\lambda) &=& \sqrt{2} \frac{\zeta k_{B}T}{h}
\Bigl [\Bigl ( 1+\frac{1}{\sqrt{2}}\Bigr )(\lambda^{2}-1)
-(\lambda^{-2}-1)\Bigr ].
\end{eqnarray}
At strong extension of a layer of chains ($\lambda\gg 1$),
Eqs. (52) and (53) show a monotonic increase in tensile
stresses with $\lambda$, but the non-entropic model
implies a higher (by a factor of $\frac{1}{2}(\sqrt{2}+1)$)
rate of growth of the extra-stress $\sigma$.

Formulas (52) and (53) demonstrate the principal
difference between the two approaches at compression.
According to the entropic elasticity theory, Eq. (52),
a finite compressive stress $\sigma_{\rm e}=-2\Sigma_{0}$
leads to the total compression of a layer of
flexible chains ($\lambda=0$).
As this conclusion contradicts basic hypotheses of
continuum mechanics, the corresponding strain energy
density, Eq. (39), should be excluded from the
consideration.
On the contrary, Eq. (53) reveals that
an infinite compressive stress is required for the
total compression of a layer of grafted chains,
in accord with the axioms of the nonlinear theory
of elasticity.

It is worth recalling that Eqs. (52) and (53) are derived
for a compressible layer of grafted chains,
which implies that they differ from conventional
relations describing uniaxial extension of an incompressible
medium (where the engineering stress is proportional
to $\lambda$, not to $\lambda^{2}$).
This discrepancy is driven by the fact that the area
of a compressible layer with the normal vector
${\bf e}_{3}$ does not change, while at uniaxial tension
of an incompressible material, this area decreases
as $\lambda^{-1}$.

It is of interest to compare predictions of the
non-entropic model with experimental data.
For this purpose, we use observations at compression
of red cell membranes by a micro-sphere tip
(the bio-membrane force probe).
A red cell ghost after a preliminary treatment was put
on a glass substrate and was compressed by a glass bead
with a micron-size radius.
The compressive force $f_{\rm c}$ was measured
simultaneously with the distance from the bead to
the substrate $z$.
The red cell membrane on the substrate is modeled as
a rigid layer, to which a compressible spectrin network
(treated as a layer of non-interacting flexible chains)
is linked by junctional complexes of short actin
filaments and other proteins.
For a description of the material, the experimental
procedure, and the method of analysis of measurements,
the reader is referred to the original publication
\cite{HRM01}.
In numerical simulation, the force $f_{\rm c}$ is found
from Eq. (53) where the stress $\sigma_{\rm ne}$ is
multiplied by the contact area $S$ (following \cite{HRM01},
the latter is assumed to be independent of $z$).
The elongation ratio $\lambda$ is connected with the
distance $z$ by the formula $\lambda=z/Z$, where $Z=205$ nm
is the initial distance (determined from the measurements
as the smallest distance at which the force $f_{\rm c}$
vanishes).
The compressive force $f_{\rm c}$ is plotted versus
distance from the substrate $z$ in Figure 1
together with the curve $f_{\rm c}(z)$ calculated at
$\sqrt{2}\zeta k_{B}TS/h =30$ pN.
Figure 1 demonstrates an acceptable agreement between
the observations and the results of numerical analysis.
We do not treat this agreement as a confirmation of
the model (the experimental data were obtained based on
a number of simplifications that may be questioned),
but rather as a demonstration of failure of the
entropic theory of rubber elasticity.

\section{Superposition of uniaxial tension and shear}

Our purpose now is to consider the influence of uniaxial
tension (compression) on shear deformation of a layer
of grafted flexible chains.
An exposition of motivation for the study
of this problem is postponed to the discussion
at the end of this section.
The analysis is focused on the non-entropic model with
the average free energy (28), whose results are compared
with those of the entropic elasticity theory.

Subsequent imposition of uniaxial tension with an
elongation ratio $\lambda$ and the deformation gradient
${\bf F}_{1}$ given by Eq. (38)
and simple shear with the deformation gradient
\[
{\bf F}_{2}={\bf e}_{1}{\bf e}_{1}
+{\bf e}_{2}{\bf e}_{2}
+{\bf e}_{3}{\bf e}_{3}
+\kappa {\bf e}_{2}{\bf e}_{3},
\]
where $\kappa$ stands for shear,
induces deformation with the deformation gradient
\begin{equation}
{\bf F}={\bf F}_{2}\cdot {\bf F}_{1}
={\bf e}_{1}{\bf e}_{1}
+{\bf e}_{2}{\bf e}_{2}
+\lambda {\bf e}_{3}{\bf e}_{3}
+\lambda \kappa {\bf e}_{2}{\bf e}_{3}.
\end{equation}
It follows from Eqs. (28) and (54) that
\begin{equation}
\Delta \Psi_{\rm ne}(\lambda,\kappa,{\bf Q})=k_{B}T
\biggl \{ \frac{3}{2b^{2}}\Bigl [ \Bigl (
(Q_{2}+\lambda\kappa Q_{3})^{2}-Q_{2}^{2}\Bigr )
+(\lambda^{2}-1)Q_{3}^{2}\Bigr ]
+\frac{b^{2}}{3Q_{3}^{2}}(\lambda^{-2}-1)\biggr \}.
\end{equation}
Inserting expressions (44) and (55) into Eq. (18)
and calculating the integral, we find that
\begin{equation}
W(\lambda,\kappa)=k_{B}T\Bigl [ C_{1}\Bigl (\lambda^{2}
(1+\kappa^{2})-1\Bigr )
+C_{2}\Bigl (\frac{1}{\lambda^{2}}-1\Bigr )\Bigr ],
\end{equation}
where the coefficients $C_{1}$ and $C_{2}$ are given
by Eq. (47).
At combined uniaxial tension and shear,
the work of external forces (per unit area of the layer
of grafted chains and unit time) reads
\[
\Pi= \Bigl ( \Sigma \frac{1}{\lambda}\frac{d\lambda}{dt}
+\Sigma_{1}\frac{d\kappa}{dt}\Bigr )h ,
\]
where $\Sigma_{1}$ denotes the (Cauchy) shear stress.
The remark that $\Sigma_{1}$ is defined per unit area in
the deformed state is important: at uniaxial tension
$\Sigma_{1}$ is less than the engineering shear stress by
a factor of $\lambda^{-1}$, while the shear displacement of
the upper boundary of the layer increases by a factor
of $\lambda$, which means that no additional
multiplier appears in the expression for the work
of shear forces.
Substitution of this expression and Eq. (40) into
Eq. (41) results in
\[
\frac{1}{\lambda} \Bigl (\Sigma-\lambda\frac{\zeta}{h}
\frac{\partial W}{\partial \lambda} \Bigr )\frac{d\lambda}{dt}
+\Bigl (\Sigma_{1}-\frac{\zeta}{h}
\frac{\partial W}{\partial \kappa}\Bigr )
\frac{d\kappa}{dt}=0.
\]
Bearing in mind that this equality is satisfied for
arbitrary functions $\lambda(t)$ and $\kappa(t)$,
we conclude that the expressions in parentheses vanish.
Equating the first expression to zero, we arrive at
Eq. (42), whereas equating the other expression to zero,
we obtain
\begin{equation}
\Sigma_{1}=\frac{\zeta}{h}
\frac{\partial W}{\partial \kappa}.
\end{equation}
It follows from Eqs. (42), (56) and (57) that
\[
\Sigma=2\frac{\zeta k_{B}T}{h} \Bigl [ C_{1}(1+\kappa^{2})
\lambda^{2}-C_{2}\lambda^{-2}\Bigr ],
\qquad
\Sigma_{1}=2\frac{\zeta k_{B}T}{h} C_{1}\kappa\lambda^{2}.
\]
Inserting expressions (49) and (50) into these relations
and splitting the longitudinal stress $\Sigma$ into the
sum of $\Sigma_{0}$ and the extra-stress $\sigma$, we find that
\begin{equation}
\sigma=\sqrt{2} \frac{\zeta k_{B}T}{h} \Bigl [
\Bigl (1+\frac{1}{\sqrt{2}}\Bigr )
(1+\kappa^{2})\lambda^{2} -\lambda^{-2}
-\frac{1}{\sqrt{2}}\Bigr ],
\qquad
\Sigma_{1}=(1+\sqrt{2})\frac{\zeta k_{B}T}{h} \kappa\lambda^{2}.
\end{equation}
Introducing the notation
\[
\bar{\sigma}=\frac{\sigma h}{\zeta k_{B}T},
\qquad
\bar{\tau}=\frac{\Sigma_{1} h}{\zeta k_{B}T},
\]
and excluding $\lambda^{2}$ from Eqs. (58), we
obtain
\[
\bar{\sigma}=\bar{\tau}\frac{1+\kappa^{2}}{\kappa}-
\frac{1+\sqrt{2}}{\sqrt{2}}\frac{\kappa}{\bar{\tau}}-1.
\]
Resolving this equation with respect to $\bar{\tau}$,
we find that
\begin{equation}
\bar{\tau}=\frac{\kappa^{2}}{2(1+\kappa^{2})}
\biggl \{(1+\bar{\sigma}) +\Bigl [ (1+\bar{\sigma})^{2}+
2\sqrt{2}(1+\sqrt{2})
\frac{1+\kappa^{2}}{\kappa}\Bigr ]^{\frac{1}{2}}
\biggr \}.
\end{equation}
The dependence of the dimensionless shear stress $\bar{\tau}$
on shear $\kappa$ is plotted in Figure 2 at various values
of the dimensionless tensile stress $\bar{\sigma}$.
This figure demonstrates that given $\bar{\sigma}$,
the shear stress monotonically increases with shear.
Given $\kappa$, the stress $\bar{\tau}$
pronouncedly grows with $\bar{\sigma}$.
At a relatively large compression
($\bar{\sigma}=-5.0$ to $-20.0$),
the function $\bar{\tau}(\kappa)$ is practically linear,
and the tangent shear modulus is rather small.
When $\bar{\sigma}$ is positive (tension), the function
$\bar{\tau}(\kappa)$ becomes strongly nonlinear: it rapidly
grows at relatively small $\kappa$ and increases as
$\sqrt{\kappa}$ at large $\kappa$.

It is of interest to compare the solution (59) with that
found by using the entropic elasticity theory.
In the latter case, an analog of Eqs. (58) is given by
\[
\sigma=2\frac{\zeta k_{B}T}{h}
\Bigl [ (1+\kappa^{2})\lambda^{2}-1\Bigr ],
\qquad
\Sigma_{1}=2\frac{\zeta k_{B}T}{h} \kappa \lambda^{2}.
\]
Introducing the dimensionless variables $\bar{\sigma}$
and $\bar{\tau}$ and excluding $\lambda^{2}$ from these
relations, we arrive at the formula
\begin{equation}
\bar{\tau}=\frac{(2+\bar{\sigma})\kappa }{1+\kappa^{2}}.
\end{equation}
Equation (60) contradicts the physical intuition:
it shows that given a tensile stress $\sigma$,
the shear stress $\Sigma_{1}$ depends on shear $\kappa$
non-monotonically: $\Sigma_{1}$ increases with $\kappa$
at small strains, reaches its maximum,
and decays to zero at large deformations.

It is necessary to provide some explanations for our
choice of shear deformation with a fixed normal stress
for the analysis.
This problem naturally arises in the study of laminar
flows of polymer melts near a rigid surface.
Experiments reveal that some chains from the melt are
grafted on the surface.
At low flow velocities and, as a consequence,
small pressures in the melt (in the model,
the pressure $p$ is equivalent to the compressive
stress $\sigma$),
a strong friction is observed between the melt and
the rigid wall which suppresses slippage of the melt
entirely.
At higher flow velocities, measurements demonstrate
a significant reduction in friction and noticeable
slippage of the melt with respect to the surface
\cite{BDeG92,deG92}, in particular, when the grafted
chains do not overlap (an ensemble of non-interacting
tethered chains).
Two reasons for this decrease in friction may be mentioned:
(i) disentanglement of polymer chains in the bulk
from grafted chains \cite{BDeG92,DM03},
and (ii) a decay in the tangent shear modulus of
the layer of grafted chains driven by the growth of
pressure.
Although the entanglement--disentanglement mechanism
may be dominant, the above analysis demonstrates
that an increase in pressure $p$ may provide substantial
contribution into the decrease of friction as well.

\section{Concluding remarks}

The nonlinear elastic behavior has been analyzed of
an individual flexible chain and an ensemble of
non-interacting chains grafted on a rigid surface.
It is demonstrated that the conventional entropic
elasticity theory leads to the conclusions that contradict
our physical intuition, which implies that its
applicability in the mechanics of polymers is questionable.
A modification of this theory is proposed, where
the average energy of a flexible chain is treated as
the governing parameter instead of the distribution
function of end-to-end vectors.
It is shown that this refinement leads to physically
plausible conclusions for an ensemble of tethered
flexible chains under uniaxial tension (compression)
and under superposition of uniaxial tension and shear.
The results of numerical simulation for the former problem
demonstrate fair agreement with observations on
compression of red cell membranes.
Our results for the latter problem reveal a novel
micro-mechanism for the decrease in friction
(with the growth of flow velocity) of polymer
melts moving near solid walls.

\newpage
\section*{List of figures}
\parindent 0 mm

{\bf Figure 1:}
The compressive force $f_{\rm c}$
versus the distance from the substrate $z$
for a red cell membrane.
Circles: experimental data \cite{HRM01}.
Solid line: results of numerical simulation.
\vspace*{2 mm}

{\bf Figure 2:}
The dimensionless shear stress $\bar{\tau}$
versus shear $\kappa$ for $\bar{\sigma}=-20.0$,
$-10.0$, $-5.0$, $-2.0$, $-1.0$, 0.0, 2.0 and 5.0,
from bottom to top, respectively.
\vspace*{120 mm}

\setlength{\unitlength}{0.75 mm}
\begin{figure}[tbh]
\begin{center}

\end{center}
\vspace*{10 mm}

\caption{}
\end{figure}

\setlength{\unitlength}{0.75 mm}
\begin{figure}[tbh]
\begin{center}
%
\end{center}
\vspace*{10 mm}

\caption{}
\end{figure}
\end{document}